# Three-Sensor 3ω-2ω Method for the Simultaneous Measurement of Thermal Conductivity and Thermal Boundary Resistance in Film-on-Substrate Heterostructures

Guang Yang, Bing-yang Cao[*]

*Key Laboratory for Thermal Science and Power Engineering of Ministry of Education, Department of Engineering Mechanics, Tsinghua University, Beijing 100084, People's Republic of China*

[*] Corresponding author: E-mail: caoby@mail.tsinghua.edu.cn

**Abstract:**

Solid heterostructures composed of substrates and epitaxial films are extensively used in advanced technologies, and their thermophysical properties fundamentally determine the performance, efficiency, reliability, and lifetime of the corresponding devices. However, an experimental method that is truly appropriate for the thermophysical property measurement of solid heterostructures is still lacking. To this end, a three-sensor 3ω-2ω method is proposed, which can simultaneously measure the thermal conductivities of the film and the substrate, along with the film-substrate thermal boundary resistance (TBR) in a single solid heterostructure without any reference samples, showing broad applicability for miscellaneous heterostructures with film thickness ranging from 100 nm to 10 μm. In this method, three parallel metal sensors with unequal width and spacing are fabricated on the sample surface, in which the two outer sensors are used as heaters, and the middle sensor is used as a detector. The respective 3ω signals of the two heaters and 2ω signal of the detector are measured, and then the thermophysical properties of the sample are fitted within 3D finite element simulations. To verify this method, two typical wide bandgap semiconductor heterojunctions, i.e., GaN on SiC (#SiC) and GaN on Si (#Si) with ~2.3 μm GaN epilayers, are measured. The thermal conductivity of the GaN film, the thermal conductivities of the SiC and Si substrates, and the GaN/substrate TBRs are derived, exhibiting good agreement with the reported values in the literature. The proposed method will provide a comprehensive solution for the thermophysical property measurements of various solid heterostructures.

# I. Introduction

Solid heterostructures are generally composed of substrates and epitaxial films, and the constituent materials include metals, semiconductors, or insulators. Thanks to their excellent and controllable performance, solid heterostructures have been widely used in power electronics, radio frequency (RF) communications, optoelectronics and photovoltaics, thermoelectricity, and other advanced technology fields[1-8]. Among them, a typical application is the GaN high electron mobility transistor (HEMT), a type of wide bandgap (WBG) semiconductor heterojunction device in power and RF electronics. By growing GaN, AlN, and AlGaN epitaxial films with a thickness of several microns on SiC, diamond, or other substrates, and integrating the wide bandgap characteristics of WBG materials and the high thermal conductivity of the substrate materials, extraordinary electrical performance (high electron mobility, high breakdown voltage, and high power gain, etc.) is finally achieved, with the improvement of the heat dissipation performance to a certain extent[9]. However, with rapid iteration of the device performance, the power density and junction temperature also rise significantly, seriously affecting the device's performance, efficiency, reliability, and lifetime[10]. Hence, researchers have extensively conducted thermal analysis and optimization design to cope with these challenges, from the near-junction regions of heterostructure devices to the external heat dissipation structures[11, 12], although this is still insufficient for the increasingly stringent and complex thermal management requirements[13]. One of the critical problems points to a high-precision experimental method that is truly appropriate for measuring the thermophysical properties of diverse solid heterostructures.

The typical experimental methods for measuring the thermophysical properties of solid heterostructures contain two main categories: optical and electrical (summarized in Table 1). The representative optical methods include the transient thermoreflectance (TTR), the steady-state thermoreflectance (SSTR), and the Raman spectroscopy (Raman), while the electrical methods mainly include the steady-state electrical method and the harmonic method (3ω method, etc.). The TTR methods include the time-domain thermoreflectance (TDTR) and frequency-domain thermoreflectance (FDTR)[14, 15]. The laser repeating frequency of TTR is usually on the order of 10 MHz, which makes the thermal penetration depth quite shallow (~O(100 nm)), resulting in a relatively low signal sensitivity to thermal boundary resistance (TBR) for the widely used heterostructures with micron-scale film thickness. To improve the measurement sensitivity of TBR, it is necessary to artificially thin the film to ~O(100 nm)[16]. Nonetheless, since the phonons transport process in the film is in the ballistic-diffusion regime[17], the thinning process inevitably makes the measured results deviate from the properties of the original heterostructures.

To overcome the limitation of the shallow thermal penetration depth, Braun et al.[18] and Song et al.[19] proposed the SSTR method. By significantly reducing the laser frequency to the order of 100 Hz and adjusting the spot size to increase the thermal penetration depth (up to 10 μm), the signal sensitivity to the TBR of heterostructures with micron-scale film thickness is indeed improved. However, the experiment results show that the method's sensitivity to the substrate thermal conductivity is still unsatisfactory, which makes it necessary to test with the help of pure substrate reference samples[19].

Raman spectroscopy generally measures local temperature according to the Stokes peak shift[20]. By measuring the internal temperature distribution of a sample and the heating power, the film and substrate thermal conductivity, as well as the TBR can be derived. The temperature measurement uncertainty of this method is ~5 K, and the vertical spatial temperature resolution is usually >1 μm. These features restrict the measurement accuracy for the film thermal conductivity and TBR[21, 22], resulting in a rather large TBR uncertainty (~10 m$^2$ K/GW)[23], which makes it difficult for Raman spectroscopy to satisfy the requirements of high-precision measurement. Additionally, Raman spectroscopy is generally capable of measuring the thermophysical properties of transparent materials[24].

Table 1. Comparison of the main features of existing experimental techniques ($T$: temperature, $\kappa_f$: film thermal conductivity, $R_I$: TBR, $\kappa_{sub}$: substrate thermal conductivity, $t_f$: film thickness)

| Category | Method | Measurability | Accuracy of $T$ | Accuracy of $R_I$ | Accuracy of $\kappa_{sub}$ | Requirements |
|---|---|---|---|---|---|---|
| Optical | TTR | $\kappa_f$, $\kappa_{sub}$, $R_I$ | High | High | Low | $t_f \sim O(100$ nm$)$ |
| | SSTR | $\kappa_f$, $R_I$ | High | High | Low | Known $\kappa_{sub}$ |
| | Raman | $R_I$ | Low | Low | Low | Transparent |
| Electrical | Steady-state | $R_I$ | High | Low | Low | / |
| | Differential 3ω | $\kappa_f$, $R_I$ | Ultra-high | Low | Low | Known $\kappa_{sub}$; serial samples varying $t_f$ |
| | Two-sensor 3ω–2ω | $\kappa_f$, $\kappa_{sub}$, $R_I$ | Ultra-high | Moderate | High | Serial samples varying $t_f$ |
| | Three-sensor 3ω–2ω | $\kappa_f$, $\kappa_{sub}$, $R_I$ | Ultra-high | High | High | / |

In terms of electrical methods, metal strips need to be fabricated on the sample surface as heaters and detectors. The steady-state electrical method heatings the samples with DC currents[25], while the harmonic methods (the most typical is the 3ω method) are sinusoidal AC heating[26]. Since the thermal penetration depth and the thermal spreading area of steady-state heating are theoretically infinite, the sensor signal of the steady-state method is more sensitive to the thermal contact between the bottom of sample and the sample carrier, and the error caused by heat radiation cannot be neglected either[27]. These factors will significantly increase the measurement error.

Among all of the kinds of harmonic methods, the most representative is the differential 3ω method[28]. However, there are three main limitations of this method. First, the method requires the measurement of a series of samples with different film thickness to obtain the film thermal conductivity, which cannot meet the requirement of deriving thermophysical properties within a single sample. Second, in this method, it is assumed that the film thermal conductivity is independent of the thickness, which means that this method only applies to amorphous films such as $SiO_2$ in principle[28]. Finally, the fitting intercept of this method includes contributions from the sensor-film TBR, the film-substrate TBR, and the substrate thermal conductivity. However, these three properties cannot be effectively distinguished, so it is necessary to substitute the substrate thermal conductivity from literature as a known value[29].

To overcome the limitations of the differential 3ω method, Hua et al. proposed a two-sensor 3ω-2ω method[24]. Likewise, this method requires the preparation of a series of heterostructure samples that only change the film thickness. A pair of parallel metal sensors with different widths are fabricated on the surface of each sample, and then the film thermal conductivity, the substrate thermal conductivity, and the film-substrate TBR are derived sequentially. Nevertheless, this method still inherits the first two limitations of the differential 3ω method. Thus, the goal of simultaneous determination of the film and substrate thermal conductivity, along with the TBR within a single sample cannot be achieved, and this method only applies to materials whose thermal conductivity is independent of thickness as well.

The so-called experimental method that is truly appropriate for measuring thermophysical properties of the solid heterostructures refers to the techniques that adapt miscellaneous heterostructures consisting of films with typical thicknesses (generally 100 nm–10 μm), and is able to simultaneously derive the target thermophysical properties (e.g., the film thermal conductivity, the substrate thermal conductivity, and the film-substrate TBR) within a single sample. In addition, it is required that no reference samples be introduced to derive part of the thermophysical properties as known values.

It is not difficult to find that a high-precision experimental method that is truly appropriate for measuring the thermophysical properties of solid heterostructures is still lacking. In this work, a three-sensor 3ω-2ω method is proposed, which is capable of simultaneously measuring the film and substrate thermal conductivity, along with the film-substrate TBR of a solid heterostructure sample without the need for any reference samples. To verify the accuracy and applicability of this method, two typical WBG heterojunction samples, i.e., GaN on SiC (#SiC) and GaN on Si (#Si), are measured. The GaN thermal conductivity, the substrate thermal conductivity, and the GaN/substrate TBR of these two samples are all within rational ranges in comparison to the literature.

## II. Experimental Method

### A. Experimental System

As shown in Fig. 1(a), this method mainly involves the measurement of the thermophysical properties of solid heterostructures composed of films and substrates. An insulating layer deposited on the film surface is required to prevent leakage and crosstalk between sensors, which is unnecessary for well-insulating film materials. Then three parallel metal sensors with different widths and spacing are fabricated on the surface by lithography, sputtering, and lift-off processes successively. The two outer sensors are heaters, among which the wider one is denoted as Heater 1 (H1), and the narrower one is denoted as Heater 2 (H2). The middle sensor is used as a Detector (D), thus forming an effective sample (i.e., device under test, DUT). The detailed sensor design guidelines are discussed in Section III.B.

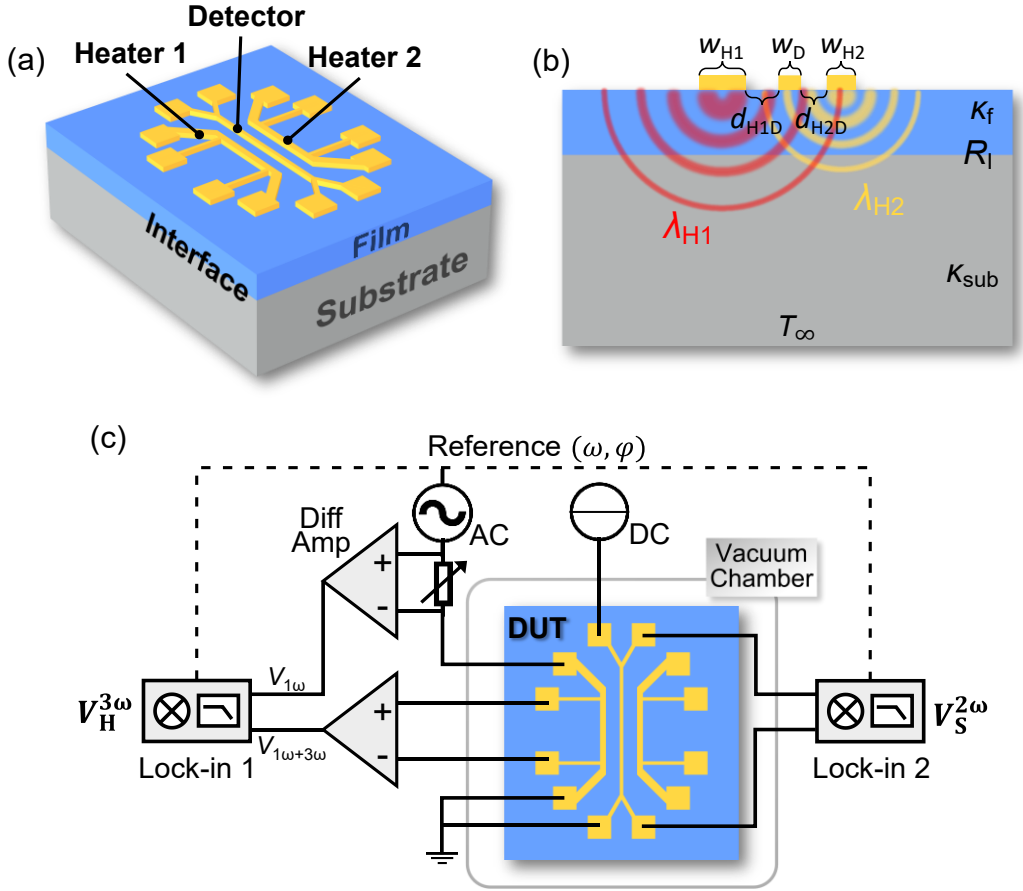

**Fig. 1.** Typical structure of a test sample (i.e., device under test, DUT) and the experiment system for the three-sensor 3ω-2ω method. (a) The heterostructure of the DUT and the three-sensor layout on the surface. (b) The heater heats the sample and the bottom temperature $T_\infty$ is controlled to be constant. The heater width and the heating frequency significantly affect the thermal penetration depth $\lambda_{H1}, \lambda_{H2}$, which therefore affects the sensitivity of the measurement signals. The divergent sensitivity to the thermophysical properties at different depths makes it possible to derive the film thermal conductivity $\kappa_f$, the substrate thermal conductivity $\kappa_{sub}$, and the TBR $R_\text{I}$ simultaneously in one test. (c) Experiment system and circuit configuration. The heaters (H1 and H2) need to be energized by AC in sequence instead of synchronously during the test, while the detector (D) is energized by DC throughout the measurement. The figure illustrates the circuit connection when H1 (or H2) and D are working synchronously.

Fig. 1(b) shows the five characteristic geometry parameters that need to be carefully designed for the three-sensor layout: H1 width $w_\text{H1}$, H2 width $w_\text{H2}$, D width $w_\text{D}$, H1-D spacing $d_\text{H1D}$, and H2-D spacing $d_\text{H2D}$. In addition, Fig. 1(b) also illustrates the relationship between the thermal penetration depth $\lambda_\text{H1}, \lambda_\text{H2}$ (where the temperature oscillation amplitude is 1/e of that at the heat source[26]) and the film thickness. By adjusting the heater width and the heating frequency, the thermal penetration depth can be directly controlled[23, 28], so the signal sensitivity of each sensor to the thermophysical properties at different depths is divergent. Therefore, it is possible to derive the film thermal conductivity $\kappa_\text{f}$, the substrate thermal conductivity $\kappa_\text{sub}$, and the film-substrate TBR $R_\text{I}$ simultaneously in a single sample.

Based on the three-sensor layout of the DUT shown in Fig. 1(a) and (b), the experiment system is set up as shown in Fig. 1(c). First, the DUT is assembled in a vacuum chamber to avoid the introduction of errors by convection and radiation and to ensure accurate temperature control. The three sensors are connected to the external circuit through a wire-bonding process, and then the two heaters are powered sequentially with an AC current source (e.g., Keithley 6221), while the detectors are powered continuously with a DC current source (e.g., Keithley 2450). The heater's 3ω signal and the detector's 2ω signal are extracted by two lock-in amplifiers (e.g., SRS SR830) respectively, with the reference frequency and phase of each lock-in amplifier provided by the AC current source.

Moreover, owing to the limitation of the lock-in amplifier's dynamic reserve, it is necessary to connect a variable resistor with the same resistance as the heater in series before the heater, which is done to prevent the lock-in amplifier from overloading. Then, the heater and the variable resistor signals are input into the lock-in amplifier under the differential mode, thereby subtracting the 1ω common-mode voltage signal[30]. In addition, since either the upper surface of the sample is generally covered with an insulating layer (i.e., amorphous $SiO_2$ or $Al_2O_3$), or the film itself is a material with good insulating performance, the 1ω voltage signal on the detector due to the leakage from the heater is negligible, and no additional common-mode voltage subtraction is required.

## B. Measurement Procedure

Consistent with the experiment system presented in Section II.A, an illustrative experimental procedure is proposed (Fig. 2), which in principle achieves the simultaneous measurement of the equivalent thermal conductivity of the insulating layer $\kappa_{\text{ins}}$, the film thermal conductivity $\kappa_{\text{f}}$, the substrate thermal conductivity $\kappa_{\text{sub}}$, and the film-substrate TBR $R_{\text{I}}$ within a single heterostructure DUT. Three factors contribute to $\kappa_{\text{ins}}$: the sensor-insulating layer TBR, the thermal resistance of the insulating layer itself, and the insulating layer-film TBR. The basic concept of this method is to determine the target thermophysical properties with the inverse problem method (fitting algorithms) based on the 3-dimensional finite element modeling (FEM).

The key operations after the preparation stage shown in Fig. 2 are the following three steps.

(1) The AC current ($I_{\text{H1}}$) is connected to the wider heater H1, while the DC current ($I_{\text{D1}}$) is connected to the detector D. Measure the heating power ($Q_1$), the 3ω signal of H1 ($V_{\text{H1}}^{3\omega}$), and the 2ω signal of D ($V_{D1}^{2\omega}$). Based on $Q_1$ and $V_{D1}^{2\omega}$, the substrate thermal conductivity $\kappa_{sub}$ is fitted by solving the inverse problem based on FEM. Specifically, optimization algorithms (Levenberg–Marquardt algorithm[31], Bayesian Optimization, etc.) are applied to minimize the deviation of the sensors' thermal responses between the lock-in amplifiers' readouts with the FEM results, and then the best-fit is drawn in correspondence to the undetermined thermophysical properties.

(2) The AC current ($I_{H2}$) is connected to the narrower heater H2, while the DC current ($I_{D2}$) is connected to the detector D. The heating power ($Q_2$), the 3ω signal of H2 ($V_{H2}^{3\omega}$), and the 2ω signal of D ($V_{D2}^{2\omega}$) are measured. Based on $Q_2$, $V_{D2}^{2\omega}$, and the $\kappa_{sub}$ obtained in the first step, the film-substrate TBR $R_I$ is fitted.

(3) Based on $Q_1$ and $V_{H1}^{3\omega}$, $Q_2$ and $V_{H2}^{3\omega}$, along with the obtained $\kappa_{sub}$ and $R_I$, the equivalent thermal conductivity of the insulating layer $\kappa_{ins}$ and the film thermal conductivity $\kappa_f$ are simultaneously fitted.

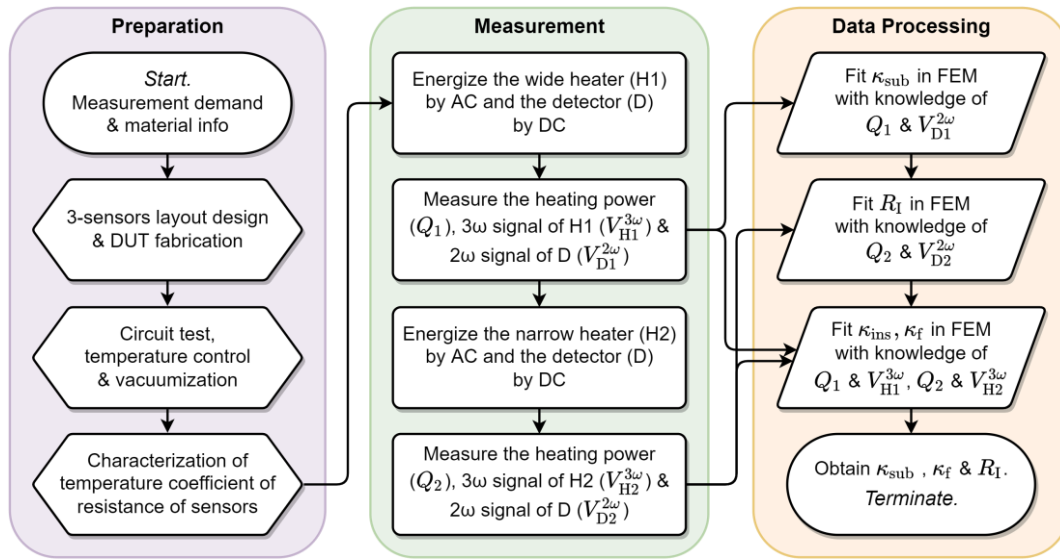

**Fig. 2.** Measurement procedure of the three-sensor 3ω-2ω method

Additionally, the amplitude of each sensor's signal is mainly utilized throughout this method due to the relatively larger error of the phase. It should be noted that the above experimental procedure is flexible. In principle, the sequential order for deriving $\kappa_{\text{ins}}$, $\kappa_{\text{f}}$, $\kappa_{\text{sub}}$ $R_{\text{I}}$, and which sensor signal is adopted are determined by the sensitivity analysis. This varies with the DUT's practical structure and material, so the derivation process of the thermophysical properties should be adapted to the actual situation, which is discussed in Section III.B.

## III. Model and Signal Analysis

### A. Model and Measurement Signals

As mentioned in Section II, the work mode of the two heaters is sequential instead of synchronous during the experiment, while the detector is working continuously during each heater heating the sample, so there are only one heater and one detector working simultaneously in this method. To clearly analyze the test signal, the DUT model composed of the two parallel sensors shown in Fig. 3 is firstly considered.

The DUT model consists of the material under test and two adjacent parallel metal sensors on the surface. It is consistent with the typical 3ω method[26] that AC current $I_{\text{H}}$ is applied to the heater,

$$I(t) = I_{\text{H}} \cos(\omega t). \tag{1}$$

The metal sensor can be regarded as a pure resistance $R^{\text{el}}$, whose inductance and capacitance can be neglected at a non-high frequency. For this condition, the heating power is the superposition of the steady-state heat source $Q^{\text{st}} (= \frac{1}{2} I_{\text{H}}^2 R_{\text{H0}}^{\text{el}})$ and the harmonic heat source $Q^{2\omega} (= \frac{1}{2} I_{\text{H}}^2 R_{\text{H0}}^{\text{el}} \cos(2\omega t))$ with a frequency of 2ω,

$$Q(t) = I(t)^2 R_{\text{H}0}^{\text{el}} = \frac{1}{2} I_{\text{H}}^2 R_{\text{H}0}^{\text{el}} [1 + \cos(2\omega t)]. \tag{2}$$

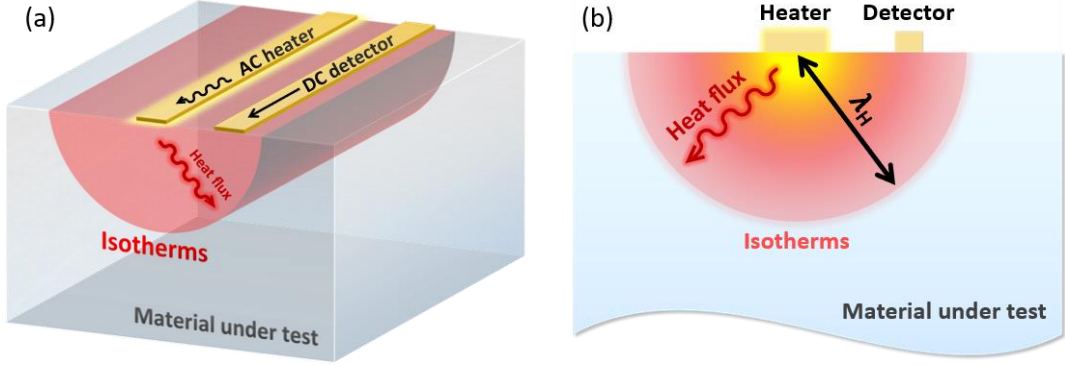

**Fig. 3.** An illustrative two-sensor DUT model. (a) Schematic of heat flux and temperature oscillation amplitude inside the sample when the heater is connected to AC and the detector in the vicinity is connected to DC. The isotherm is the isosurface of temperature amplitude $\theta^{2\omega}$, excluding the steady-state temperature rise component $\theta^{\text{st}}$. The DC current is usually small, so the heat flux generated by the detector is negligible. (b) Schematic of the thermal penetration depth $\lambda_{\text{H}}$ of the harmonic heat source $Q^{2\omega}$.

$R_{\text{H}0}^{\text{el}}$ represents the heater's electrical resistance at the reference temperature. In addition, the DUT shown in Fig. 3 (and Fig. 1(a)) can be regarded as a linear thermal impedance device[28, 32, 33]. Thus, the thermal response signal $\theta$ (i.e., the temperature change) can be directly decomposed into two components corresponding to the heat source term in Eq. (2): the steady-state temperature rise $\theta^{\text{st}}$ associated with the steady-state heat source $Q^{\text{st}}$, and the temperature oscillation $\theta^{2\omega}$ associated with the harmonic heat source $Q^{2\omega}$ [28]. Then, the thermal responses of the heater and detector can be denoted as

$$\begin{cases} \theta_{\text{H}} = \theta_{\text{H}}^{\text{st}} + \theta_{\text{H}}^{2\omega} \cos(2\omega t + \varphi), & (3a) \\ \theta_{\text{D}} = \theta_{\text{D}}^{\text{st}} + \theta_{\text{D}}^{2\omega} \cos(2\omega t + \psi). & (3b) \end{cases}$$

The phase $\varphi$ of the heater thermal response reflects the time delay caused by the heat capacity of the heater itself and the sample, while the $\psi$ of the detector reflects the influence of the heat capacity and the distance between the detector and the heater[34]. In addition, when the temperature change is small, there is a good linear relationship between the metal electrical resistance and the temperature, which can be characterized by the temperature coefficient of the electrical resistance (TCR, denoted by $\beta = \frac{1}{R_0^{el}}\frac{\partial R^{el}}{\partial T}$, $R^{el} = R_0^{el}(1+\beta\theta)$)[32, 35, 36]. Hence, the resistance changes of the heater and the detector due to temperature changes are denoted as follows:

$$\begin{cases} R_H^{el} = R_{H0}^{el}[1 + \beta_H\theta_H^{st} + \beta_H\theta_H^{2\omega}\cos(2\omega t + \varphi)], & (4a) \\ R_D^{el} = R_{S0}^{el}[1 + \beta_D\theta_D^{st} + \beta_D\theta_D^{2\omega}\cos(2\omega t + \psi)]. & (4b) \end{cases}$$

A sinusoidal AC current at 1ω frequency and the resistance change at 2ω frequency result in a 3ω component in the voltage signal on the heater[37],

$$V_H(t) = I_H R_{H0}^{el}\begin{bmatrix} (1 + \beta_H\theta_H^{st})\cos(\omega t) + \frac{1}{2}\beta_H\theta_H^{2\omega}\cos(\omega t + \varphi) \\ +\frac{1}{2}\beta_H\theta_H^{2\omega}\cos(3\omega t + \varphi) \end{bmatrix}. \quad (5)$$

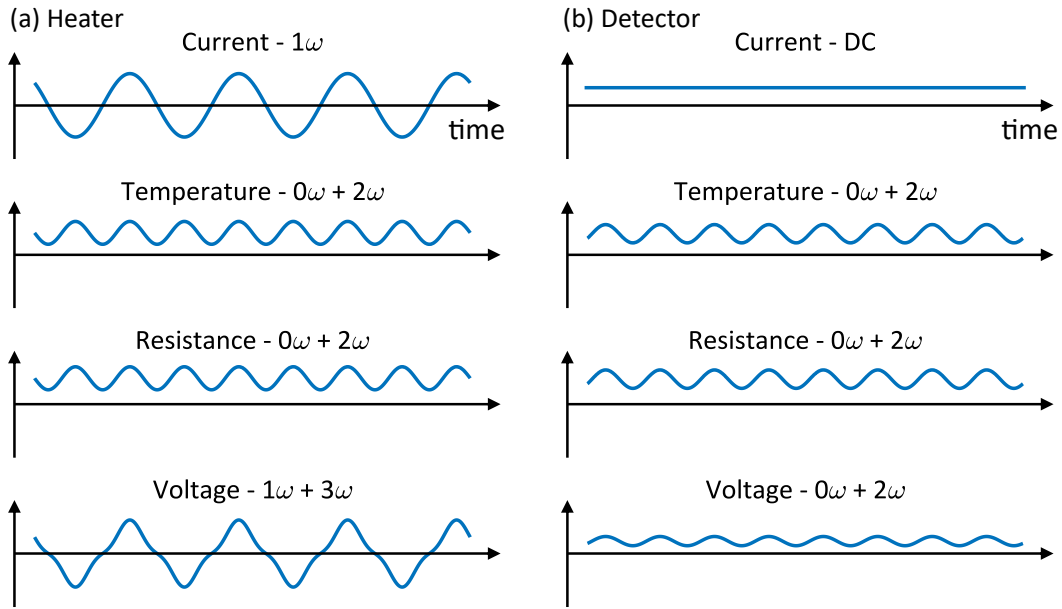

**Fig. 4.** Schematic of waveforms of the applied current, temperature oscillation, resistance change, and voltage response on each sensor. (a) Heater, (b) detector.

Meanwhile, a constant DC current $I_D$ is connected to the detector. The constant DC current and the resistance change of the 2ω frequency cause the voltage signal on the detector to contain a 2ω component,

$$V_D(t) = I_D R_{s0}^{el}\left[1 + \beta_D \theta_D^{st} + \beta_D \theta_D^{2\omega} \cos(2\omega t + \psi)\right]. \tag{6}$$

It should be noted that the $I_D$ also generates an additional steady-state temperature rise $\theta'$ on the sample, the heater, and the detector. However, since $I_D$ is generally small, and the steady-state temperature rise $\theta'$ does not affect the 3ω signal of the heater and the 2ω signal of the detector, the $\theta'$ can therefore be neglected in principle. Fig. 4 summarizes and laterally compares the current, temperature oscillation, resistance change, and voltage response waveforms of the heater and the detector.

It is obvious from Eqs. (5) and (6) that the 3ω component $V_H^{3\omega}$ of the heater voltage signal and the 2ω component $V_D^{2\omega}$ of the detector voltage signal are both linearly related to the temperature oscillation $\theta^{2\omega}$ [26, 35, 38], i.e.,

$$\begin{cases} \theta_H^{2\omega} = \dfrac{2\, V_H^{3\omega,rms}}{I_H^{rms} R_{H0}^{el} \beta_H}, & (7a) \\ \theta_D^{2\omega} = \dfrac{\sqrt{2}\, V_D^{2\omega,rms}}{I_D R_{s0}^{el} \beta_D}. & (7b) \end{cases}$$

Therefore, by detecting the RMS values of $V_H^{3\omega}$ and $V_D^{2\omega}$ through the lock-in amplifiers, the temperature oscillation on each sensor can be directly obtained using Eq. (7), whose amplitude and phase contain the thermophysical properties of the DUT.

### B. Sensitivity Analysis

Based on the dimensionless sensitivity of the measurement signal (Eq. 8), the feasibility of this method for deriving the target thermophysical properties, and the rationality of the three-sensor layout applied on the DUT shown in Fig. 1 are explained.

$$S_i = \left| \frac{\partial \ln(\theta^{2\omega})}{\partial \ln(p_i)} \right| \approx \left| \frac{\Delta \theta^{2\omega}/\theta^{2\omega}}{\Delta p_i/p_i} \right|, \tag{8}$$

where $\theta^{2\omega}$ is the amplitude of the temperature oscillation of each sensor ($\theta_H^{2\omega}$ for the heater, $\theta_D^{2\omega}$ for the detector), and $p_i$ represents the four undetermined thermophysical properties of the DUT: the equivalent thermal conductivity of the insulating layer $\kappa_{ins}$, the film thermal conductivity $\kappa_f$, the substrate thermal conductivity $\kappa_{sub}$, and the film-substrate TBR $R_I$. The $\kappa_{ins}$ includes the contributions from three factors: the metal sensor-insulating layer TBR $R_I^{m\text{-}ins}$, the thermal resistance of the insulating layer itself $\frac{t_{ins}}{\kappa_i}$, and the insulating layer-film TBR $R_I^{ins\text{-}f}$, which can be quantized with

$$\kappa_{ins} = \frac{t_{ins}}{R'} = \frac{t_{ins}}{R_I^{m\text{-}ins} + \frac{t_{ins}}{\kappa_i} + R_I^{ins\text{-}f}} = \left( \frac{R_I^{m\text{-}ins}}{t_{ins}} + \frac{1}{\kappa_i} + \frac{R_I^{ins\text{-}f}}{t_{ins}} \right)^{-1}. \tag{9}$$

Since the insulating layer is very thin (generally ⩽50 nm to meet insulation requirements) compared to the heaters with widths of several microns, the heat conduction in the insulating layer can be regarded as a one-dimensional process[28]. In this condition, the three parts $R_I^{m\text{-}ins}$, $\frac{t_{ins}}{\kappa_i}$, $R_I^{ins\text{-}f}$ can be regarded as being in series. With this premise, the contributions of $R_I^{m\text{-}ins}$ and $R_I^{ins\text{-}f}$ to the total thermal resistance of the insulating layer can be simply incorporated into the equivalent thermal conductivity of the insulating layer $\kappa_{ins}$. Thus, Eq. (9) is reasonable.

Herein, a typical heterostructure consisting of a GaN epitaxial film and a SiC substrate (GaN on SiC) is analyzed, and benchmark parameters of the model are set as follows: insulating layer thickness $t_{ins} = 50$ nm, GaN film thickness $t_f = 2.5$ μm, substrate thickness $t_{sub} = 350$ μm, equivalent thermal conductivity of the insulating layer $\kappa_{ins} = 0.6$ W/m K, film thermal conductivity $\kappa_f = $ 130 W/m K, substrate thermal conductivity $\kappa_{sub} = 370$ W/m K, film-substrate TBR $R_I = 40$ m² K/GW, heating current frequency $f_H = 600$ Hz, and the length of each sensor $L = 600$ μm. Additionally, each material's heat capacity and density are taken from the values in the literature[39-41]. The calculations are performed in 3D FEM.

The analyses mainly focus on three characteristic geometric parameters: the heater width $w_H$, the heater-detector spacing $d_{HD}$, and the detector width $w_D$. These geometric parameters determine the three-sensor layout. By adjusting these three parameters, the sensitivities of the heater signal $\theta_H^{2\omega}$ and the detector signal $\theta_D^{2\omega}$ to each thermophysical property are calculated, and the results are shown in Fig. 5.

First, the influence of the heater width $w_H$ on the signal sensitivity is analyzed. As shown in Fig. 5(a), in region I, the heater is rather narrow, and its signal $\theta_H^{2\omega}$ is more sensitive to the thermophysical properties of the insulating layer on the surface, since the thermal penetration depth is positively correlated to the heat source width under the same heating frequency[23]. Thus, the insulating layer mainly contributes to the heater's thermal response in this condition. As $w_H$ increases, the thermal penetration depth increases significantly, and the interior thermal conduction process within the insulating layer and the epitaxial film is approximately one-dimensional, while the sensitivity of $\theta_H^{2\omega}$ to $\kappa_{ins}$ decreases significantly (Region II). Once $\kappa_{sub}$ and $R_I$ are known, the simultaneous determination of $\kappa_{ins}$ and $\kappa_f$ is achievable by combining the heater signals $\theta_H^{2\omega}$ of both region I and region II. Fortunately, it will be seen in Fig. 5(b) and (c) that $\kappa_{sub}$ and $R_I$ can be separately derived from the information of region III and region IV respectively, so the information of the region I and region II can be combined to determine $\kappa_{ins}$ and $\kappa_f$ simultaneously. In addition, it should be noted that too narrow heater width $w_H$ is not recommended. When $w_H$ approaches the phonon mean free path (MFP) of the film material, the thermal response of the heater will be significantly higher than the FEM simulation based on Fourier's law due to phonon ballistic transport (Knudsen effect)[17, 38, 42, 43]. In this case, numerical simulations regarding non-Fourier transport must be adopted.

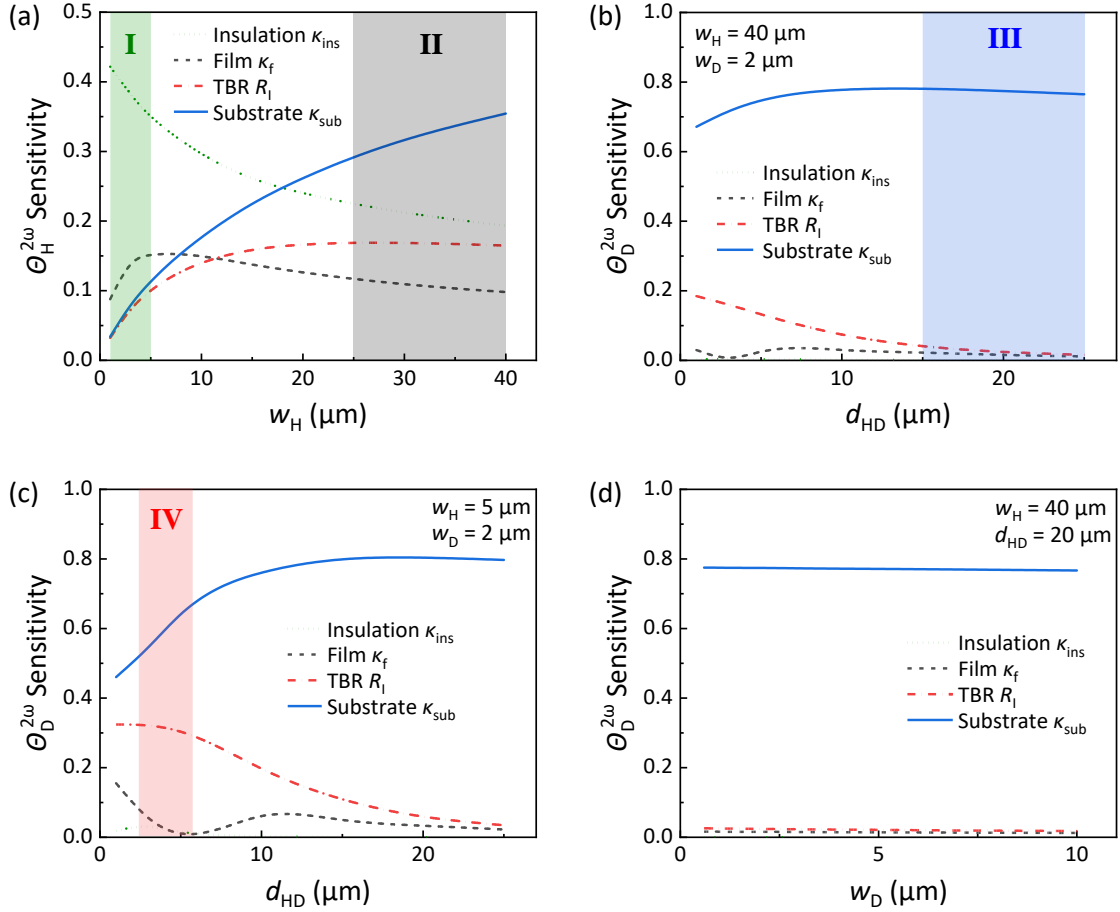

**Fig. 5.** Sensitivity analysis of the four thermophysical properties for a typical GaN on SiC heterostructure. (a) The sensitivity of the heater signal $\theta_H^{2\omega}$ to the four thermophysical properties varies with the heater width $w_H$. (b) When $w_H = 40$ μm, $w_D = 2$ μm, the sensitivity of the detector signal $\theta_D^{2\omega}$ to the four thermophysical properties varies with the heater-detector spacing $d_{HD}$. (c) When $w_H = 5$ μm, $w_D = 2$ μm, the sensitivity of $\theta_D^{2\omega}$ to the four thermophysical properties varies with $d_{HD}$. (d) The sensitivity of $\theta_D^{2\omega}$ to the four

thermophysical properties is independent of the detector width $w_D$. Region I represents the regime where the heater signal is sensitive to the equivalent thermal conductivity of the insulating layer $\kappa_{ins}$. The information of both region I and region II can be utilized to achieve the simultaneous determination of $\kappa_{ins}$ and $\kappa_f$. Region III represents the design window in which the detector signal is only sensitive to the substrate thermal conductivity $\kappa_{sub}$. Region IV represents the design window in which the detector signal is sensitive to the film-substrate TBR $R_I$ but insensitive to $\kappa_{ins}$ and $\kappa_f$. These four regions can be used to derive the four thermophysical properties respectively.

Next, the effect of the heater-detector spacing $d_{HD}$ on the signal sensitivity is analyzed. Fig. 5(b) shows the effect of adjusting $d_{HD}$ on the sensitivity of the detector signal $\theta_D^{2\omega}$. It is obvious that when $d_{HD}$ is large, signal $\theta_D^{2\omega}$ is only sensitive to $\kappa_{sub}$ (region III). As $d_{HD}$ decreases, the sensitivity of $\theta_D^{2\omega}$ to $R_I$ increases gradually, while the sensitivity to $\kappa_f$ and $\kappa_{ins}$ remains low. Hence, region III is an ideal window for deriving the substrate thermal conductivity $\kappa_{sub}$.

Fig. 5(c) shows the variation in the sensitivity of $\theta_D^{2\omega}$ with $d_{HD}$ when the heater width is narrower ($w_H = 5$ μm). As $d_{HD}$ becomes smaller, the sensitivity of $\theta_D^{2\omega}$ to $R_I$ is greatly improved (region IV) compared to Fig. 5(b). Nevertheless, when $d_{HD}$ is further reduced below 1 μm, the sensitivity of $\theta_D^{2\omega}$ to $\kappa_f$ increases rapidly as well, which is inconducive to the separate extraction of $R_I$. Moreover, to avoid excessive signal crosstalk between the heater and the detector which affects the measurement accuracy significantly, the spacing $d_{HD}$ cannot be too small (usually $d_{HD} > 500$ nm). Therefore, to accurately extract the film-substrate TBR $R_I$ and avoid sensor crosstalk, it is necessary to design the heater-detector spacing $d_{HD}$ finely within the window of region IV.

Finally, the influence of the detector width $w_D$ is also analyzed (Fig. 5(d)). Obviously, the sensitivity of signal $\theta_D^{2\omega}$ is almost independent of the detector width $w_D$. According to Eq. (7), in order to obtain a more significant $\theta_D^{2\omega}$ with a smaller detection current $I_D$ while reducing the steady-state Joule heating generated by $I_D$, it is advantageous to increase the detector's electrical resistance $R_{s0}^{el}$ by narrowing the detector. Furthermore, since a constant DC is connected to the detector, any changes in the heater-detector spacing $d_{HD}$ and the detector width $w_D$ will not affect the sensitivity of $\theta_H^{2\omega}$ to each thermophysical property in principle.

According to the above analysis, taking advantage of the significant divergent sensitivity of each sensor's thermal response signal to the underdetermined thermophysical properties, the sensor's geometric parameters corresponding to the four regions (Fig. 5) can be determined. Thus, using a set of experimental data, the equivalent thermal conductivity of the insulating layer $\kappa_{ins}$, the film thermal conductivity $\kappa_f$, the substrate thermal conductivity $\kappa_{sub}$, and the film-substrate TBR $R_I$ of the sample can be fitted simultaneously based on the inverse problem method.

Specifically, corresponding to the four regions shown in Fig. 5, the three-sensor layout design criteria of the DUT shown in Fig. 1 can be consolidated. First, corresponding to region III, a wide heater H1 is arranged on the upper surface of the sample, and a detector D1 is located at a relatively long distance from H1. This group of sensors is used to measure $\kappa_{sub}$. Then, corresponding to region IV, a narrow heater H2 is arranged, and a detector D2 is in the immediate vicinity. This group of sensors is used to derive $R_I$ based on the $\kappa_{sub}$ measured in the first step. Further, corresponding to regions I and II, the signals of the wide heater H1 and the narrow heater H2 are synthesized to derive $\kappa_{ins}$ and $\kappa_f$. It should be noted that D1 and D2 can be merged into a single detector D to simplify the sensor layout without affecting the measurement accuracy. Finally, based on the sensitivity analysis, the DUT's three-sensor layout design shown in Section II and the experimental procedure shown in Fig. 2 can be directly given.

Additionally, the generality of the above design criteria is validated. The sensitivity analysis of other three kinds of typical heterostructures, i.e., GaN on Si epitaxial heterojunction (GaN on Si), β-$Ga_2O_3$ on SiC epitaxial heterojunction ($Ga_2O_3$ on SiC), and amorphous $Al_2O_3$ on Si epitaxial heterojunction (a$Al_2O_3$ on Si), are conducted (see Supplementary Material). For GaN on Si, the parameters are identical to those of GaN on SiC, except that the Si thermal conductivity is set to 150 W/m K. For $Ga_2O_3$ on SiC, the thickness of the β-$Ga_2O_3$ layer is set to 6.5 μm[19], while the β-$Ga_2O_3$ thermal conductivity and heat capacity are obtained from the literature[16, 44], and the other parameters are consistent with GaN on SiC. For a$Al_2O_3$ on Si, the thickness of the a$Al_2O_3$ layer is set to 200 nm[24], while the a$Al_2O_3$ thermal conductivity and the heat capacity are obtained from the literature[24], and the other parameters are consistent with GaN on Si. Although there may be a few divergences from the four regions shown in Fig. 5, it is still feasible to derive the undetermined thermophysical properties separately by adjusting the experimental procedure according to the specific sensitivity conditions, which proves the good versatility of this method for various heterostructures and film thicknesses (100 nm–10 μm).

## IV. Results and Discussions

To verify the validity and the accuracy of the proposed three-sensor 3ω-2ω method, two kinds of typical heterojunctions, i.e., GaN on SiC (#SiC, 4H-SiC substrate with GaN film) and GaN on Si (#Si, undoped Si substrate with GaN film), are measured at 300 K. Both samples are prepared using the Metal Organic Chemical Vapor Deposition (MOCVD) process with annealing. For the sample #SiC, an AlN nucleation layer with a ~40 nm thickness is first grown on a 90 μm 4H-SiC substrate to buffer the stress caused by the lattice and thermal expansion mismatch between the GaN film and the substrate, improving the epitaxial quality of GaN. Then a ~2.3 μm GaN epilayer is sequentially grown on AlN to form a GaN/AlN/SiC "sandwich" structure. For the sample #Si, a ~300 nm AlN nucleation layer is grown on a 95 μm undoped Si(111) substrate, and then a ~2.3 μm GaN epitaxial layer is grown on AlN to form a GaN/AlN/Si "sandwich" structure. The significant difference in the nucleation layer thickness is attributed to the much more serious lattice and thermal expansion coefficient mismatch between Si and GaN than that of SiC[45, 46]. To improve the lattice quality of the GaN epilayer of the sample #Si, the AlN thickness needs to be increased. Owing to the existence of the AlN nucleation layer, the film-substrate TBR $R_I$ measured in this study is actually the superposition of three factors, including the GaN/AlN TBR, the thermal resistance of the AlN nucleation layer, and the AlN/substrate TBR.

To avoid the leakage current and suppress the crosstalk between sensors, a SiO₂ insulating layer of ~40 nm is deposited on the GaN layer with the Plasma Enhanced Chemical Vapor Deposition (PECVD) process, and then a complete heterostructure sample is formed. The layered structures of both samples are characterized by transmission electron microscope (TEM), and are shown in Fig. 6(a) and (b). It is obvious that there are more dislocation cores near the AlN/Si interface (black dots shown in Fig. 6(b)) of #Si, which ascribes to the more serious mismatch between Si and AlN. Then the corresponding three-sensor layouts are fabricated on the SiO₂ upper surface of the two samples via lithography, sputtering, and lift-off processes successively, and complete DUTs are finally formed[24]. The material of the sensors in this study is 90 nm Au/10 nm Cr (Cr is the adhesion layer), and the actual sensor morphology is shown in Fig. 6(c) and (d). The three-sensor layout is designed based on the sensitivity analysis demonstrated in Section III.B. For both samples, the specific sensor dimensions are measured with scanning electron microscope (SEM), respectively (Table 2). The measurement details are shown in Supplementary Material.

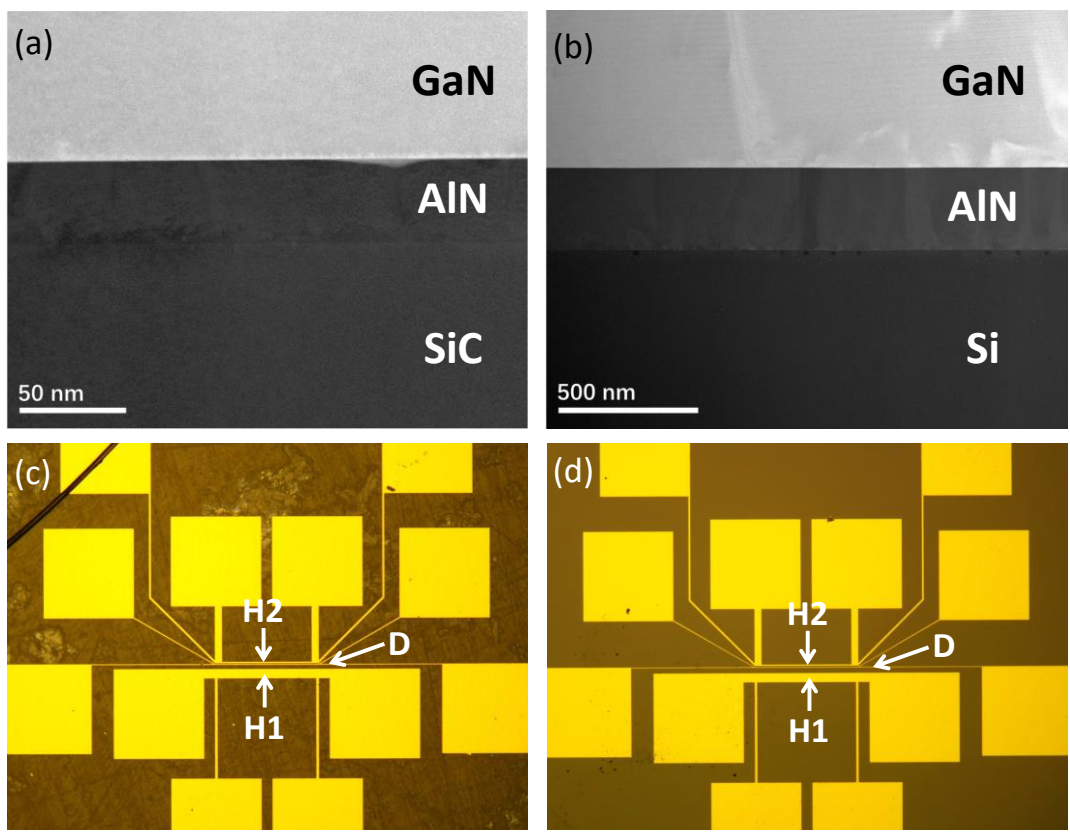

**Fig. 6.** The "sandwich" structure and the three-sensor morphology of each sample. (a) GaN/AlN/SiC "sandwich" structure of the sample #SiC, (b) GaN/AlN/Si "sandwich" structure of the sample #Si. It is obvious that there are more dislocation cores (black dots) near the AlN/Si interface of #Si. Actual three-sensor layout of (c) the sample #SiC and (d) the sample #Si.

According to the experimental procedure illustrated in Fig. 2, after the DUT fabrication and the experimental circuit test, each sensor's temperature coefficient of electrical resistance (TCR, denoted by $\beta$) needs to be calibrated first. The calibration results show that the TCRs for each sensor on the two samples are close (~$1.8\times10^{-3}$/K near 300 K, see Supplementary Material), which is ascribed to each metal sensor consisting of the same material.

Table 2. Specific dimensions of each sensor on both samples (units: μm)

| Sample | $w_{H1}$ | $w_{H2}$ | $w_D$ | $d_{H1D}$ | $d_{H2D}$ |
|---|---|---|---|---|---|
| #SiC | 42.50 | 7.741 | 4.689 | 17.19 | 2.233 |
| #Si | 42.95 | 8.113 | 4.987 | 21.81 | 2.754 |

Next, AC currents are sequentially applied to the heaters H1 and H2 of each sample, and a constant DC current is applied to the detector D while H1 or H2 is heating. Then the heating power is adjusted and the trend of the thermal response $\theta^{2\omega}$ of each sensor is recorded. The results are shown in Fig. 7, in which the scattered points are the raw data, and the error bars of data points are so small that they are covered by the points. The straight lines correspond to the best-fit thermophysical properties combinations ($\kappa_{ins}$, $\kappa_f$, $\kappa_{sub}$, and $R_I$), which are obtained by solving the inverse problem based on the optimization algorithm (i.e., the Levenberg-Marquardt algorithm[31]) within the 3D FEM simulations. It is obvious that there is an excellent linear relationship between the thermal response $\theta^{2\omega}$ of each sensor and the heating power, which verifies the fact that both samples can be regarded as linear thermal impedance devices and validates the high precision of the measurement.

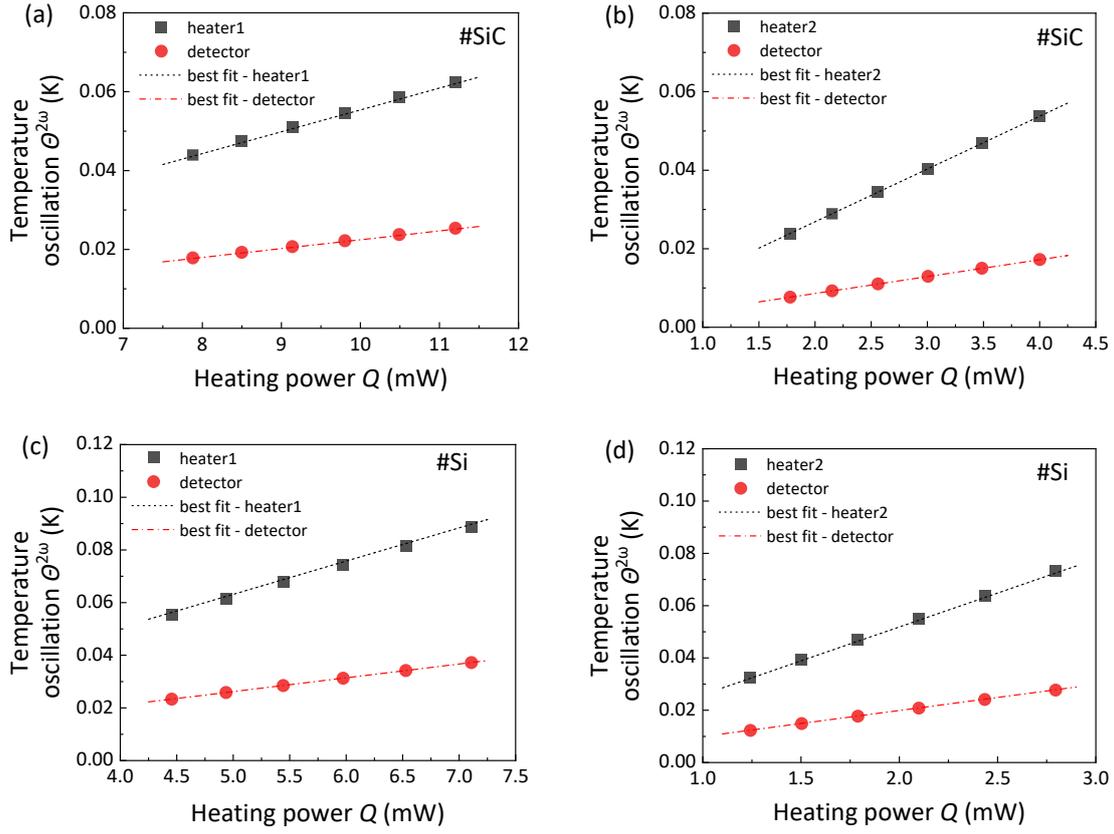

**Fig. 7.** Measurements of the two samples. The sample #SiC: (a) thermal response signals of H1 and D during H1 heating, (b) thermal response signals of H2 and D during H2 heating. The sample #Si: (c) thermal response signals of H1 and D during H1 heating, (d) thermal response signals of H2 and D during H2 heating. The scattered points are the raw data, and the straight lines are the best-fit lines corresponding to the optimal thermophysical property combinations obtained by solving the inverse problem based on the FEM simulation.

The final measurement results at room temperature (300 K) are shown in Table 3, and the error analysis is in Supplementary Material. For the sample #SiC, the GaN film thermal conductivity is 167.9 ± 7.4 W/m K, and the SiC substrate thermal conductivity is 389.1 ± 9.6 W/m K. For #Si, the thermal conductivity of the GaN film is 165.4 ± 10.2 W/m K, and the Si substrate thermal conductivity is 147.2 ± 3.7 W/m K. The thermal conductivities of the GaN film for both samples are close, and all the thermal conductivity measurements discussed above are in good agreement with the reported data in the literature[20, 39, 45, 47-49].

Table 3. Measured thermophysical properties of the two samples at 300 K

| Sample | $\kappa_{GaN}$ (W/m K) | $\kappa_{sub}$ (W/m K) | $R_I$ (m$^2$ K/GW) |
|---|---|---|---|
| #SiC | 167.9 ± 7.4 | 389.1 ± 9.6 | 5.1 ± 1.0 |
| #Si | 165.4 ± 10.2 | 147.2 ± 3.7 | 11.7 ± 2.1 |

For thermal boundary resistance, the GaN/SiC TBR of the sample #SiC is measured to be 5.1 ± 1.0 m$^2$ K/GW, while the GaN/Si TBR of the sample #Si is 11.7 ± 2.1 m$^2$ K/GW. The GaN/SiC TBR is relatively lower than that for GaN/Si, which may be ascribed to two factors. First, the matching of the lattice constant and the thermal expansion coefficient between SiC, GaN, and AlN is better than that of the Si substrate, which results in a higher lattice quality near the AlN/SiC interface of the #SiC, with fewer lattice defects and dislocation cores (as shown in Fig. 6(a) and (b)). Therefore, the scattering rate of phonons near the nucleation layer in the #SiC is smaller than that of the #Si in principle. Second, the defect density in the AlN nucleation layer is extremely high (atomic vacancy density >10$^{21}$ cm$^{-3}$)[20, 45, 46, 50], and the thickness of AlN is thin, so the phonons scattering by the defect and boundary in this layer are significant, resulting in the thermal conductivity of AlN (generally <25 W/m K[48]) being much lower than the bulk value. Although the increase of the AlN layer thickness can reduce phonon boundary scattering rate and somewhat improve the thermal conductivity, it is still not enough to offset the increase in total thermal resistance caused by the increasing thickness. Thus, the thicker AlN layer is responsible for the larger TBR of the sample #Si.

**Table 4.** Representative TBR measurements in literature (300 K) (NL: nucleation layer)

| Substrate | Study | Method | NL | Growth | $R_I$ (m$^2$ K/GW) |
|---|---|---|---|---|---|
| SiC | This work | 3ω-2ω | 40 nm AlN | MOCVD | 5.1 ± 1.0 |
| | Mu et al.[49] | TDTR | No | SAB<br>SAB and anneal | 5.9 + 0.7 /– 0.5<br>4.4 + 1.3 /– 0.8 |
| | Ziade et al.[39] | FDTR | No | MBE | 4.3 + 0.5 /– 0.4 |
| | Cho et al.[45] | TDTR | 36 nm AlN | MOCVD | 5.3 ± 1.3 |
| | Manoi et al.[51] | Raman | 70 nm AlN | MOCVD | 20.1 ± 5.0 |
| Si | This work | 3ω-2ω | 300 nm AlN | MOCVD | 11.7 ± 2.1 |
| | Bougher et al.[48] | TDTR | 100 nm AlN | MOCVD | 7.0 ± 1.7 |
| | Cho et al.[45] | TDTR | 38 nm AlN | MBE | 7.8 ± 1.2 |

The TBR measurement results in this work are compared with representative studies in the literature, as shown in Table 4 and Fig. 8. For the sample #SiC, the TBR result in this study is quite consistent with that of the MOCVD sample with a ~36 nm AlN nucleation layer prepared by Cho et al.[45]. Moreover, our result is slightly higher than those of the annealed samples with no nucleation layer prepared by Mu et al.[49] using a surface activated bonding process (SAB(A)), and the samples with no nucleation layer prepared by Ziade et al.[39] via the molecular beam epitaxy (MBE) process. In addition, the TBR of the #SiC is slightly lower than that of the unannealed samples (SAB) prepared by Mu et al.[49], and is significantly lower than that of the MOCVD sample with a ~70 nm AlN nucleation layer prepared by Manoi et al.[51].

Among the aforementioned fabrication processes, the MBE process generally produces heterostructures with the highest quality interface characterized by atomic level flatness and few defects.

Moreover, since the temperature in the MBE process is relatively low (700 °C–800 °C), the influence of the thermal expansion coefficient mismatch is small, and the atom diffusion near the interface is weak[52-54]. Therefore, for materials with good lattice matching such as GaN and SiC, the MBE process can directly grow the GaN epilayer on a SiC substrate without introducing additional AlN transition layers. Hence, the TBR of heterostructures prepared with MBE is generally low.

In terms of the SAB technique, the bonding surface of the GaN and SiC needs to be bombarded with Ar ion beams to activate the surface before the bonding process, leading to the formation of a nanometer-thick amorphous layer near the bombarded surface, and then the GaN and SiC are bonded together. After the bonding process is completed, an amorphous layer with a thickness of 3–5 nm remains near the interface[49]. Owing to the extremely low thermal conductivity of the amorphous materials (generally <2 W/m K at room temperature), the TBR is generally higher than that of the MBE samples. To improve the interface quality of the SAB samples, post-processing such as high-temperature annealing is beneficial. After annealing, the amorphous areas near the interface are significantly reduced, replaced by polycrystalline structures and dislocations after high-temperature recrystallization[49]. Thus, the interface morphology and the TBR after annealing are close to those of the MBE samples.

MOCVD is a mature process that is widely used in the fabrication of semiconductor heterostructures, and the temperature in the MOCVD process is higher (up to 1100 °C)[55] compared with MBE and SAB. Accordingly, in addition to lattice matching, the thermal expansion coefficient mismatch is not negligible, and an additional nucleation layer between the substrate and the epilayer is necessary to buffer the stress during epitaxial growth. Although the nucleation layer is helpful, the lattice defects and dislocations in the nucleation layer are still dense, while the atom diffusion and amorphous areas near the interface are still obvious. Hence, the TBR of the MOCVD sample is generally higher than that of the MBE sample and the annealed SAB sample. In the previous studies, Cho et al.'s sample structure and growth process[45] are very similar to those in this study, and the measured TBR value is likewise close to that of this study. Furthermore, MOCVD is widely used in large-scale production, and the growth rate is adjusted in a wide range by controlling the gas flow rate of the reaction source, which inevitably results in large deviations in the growth quality of MOCVD samples provided by different suppliers[51]. Hence, it is understandable that the TBR result of the MOCVD sample prepared by Manoi et al.[51] is quite different from others.

For the sample #Si, the TBR result is higher than those of the MBE samples with a ~36 nm AlN nucleation layer fabricated by Cho et al.[45], and the MOCVD samples with a ~100 nm AlN nucleation layer prepared by Bougher et al.[48]. This deviation is mainly ascribed to the much thicker AlN nucleation layer of the sample #Si, which increases the equivalent thermal boundary resistance due to its reduced thermal conductivity[48]. If the excess AlN thickness in the sample #Si is subtracted, a TBR value closer to Cho et al.'s[45] and Bougher et al.'s[48] can be obtained.

In addition, the TBR results are compared to the DMM model predictions[23]. As shown in Fig. 8, the GaN/SiC TBR predicted by the DMM model is ~1.1 $m^2$ K/GW, with the GaN/Si TBR being ~0.8 $m^2$ K/GW, which are both significantly lower than the experimental value. Generally, the DMM model underestimates the TBR significantly[23, 24], since DMM only considers the mismatch of the phonon DOS and group velocity inside the material on both sides of the interface, ignoring the anharmonic effect, the interfacial localized phonon modes, and the scattering effects of dense defects near the interface. Moreover, the additional thermal resistance corresponding to the nucleation layer introduced in practical heterostructures is absent in DMM models as well.

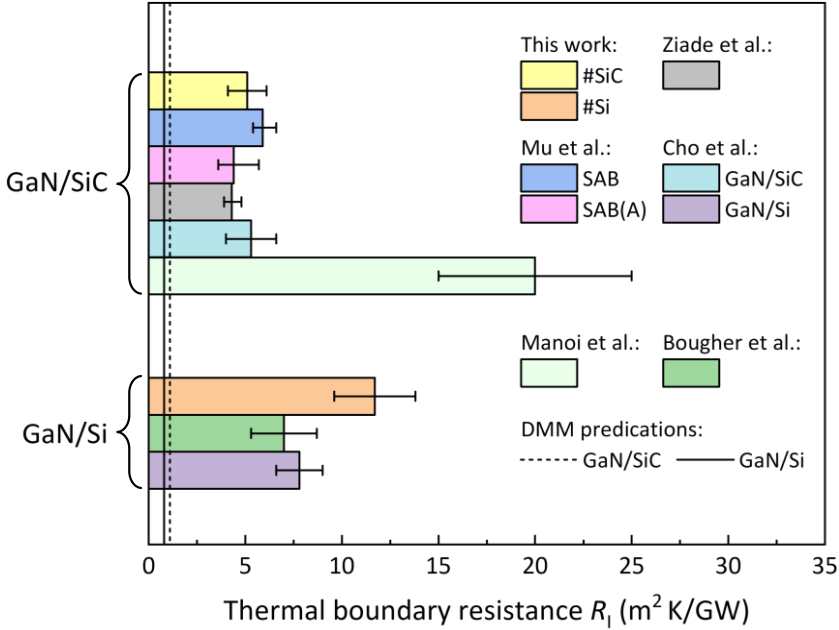

**Fig. 8.** Comparison of TBR results in this study with representative studies in the literature (300 K).

In summary, the measurement results given by the proposed three-sensor 3ω-2ω method are all consistent with the existing representative studies in the literature, which verifies the method's accuracy, reliability, and applicability to the thermophysical property measurements of typical heterostructures with film thicknesses of several microns.

## V. Conclusions

A three-sensor 3ω-2ω method is proposed in this study, which can simultaneously measure the thermal conductivities of the film and the substrate, and the film-substrate TBR within a single heterostructure sample, exhibiting broad applicability for heterostructures with film thicknesses ranging from 100 nm to 10 μm. In this method, three parallel metal sensors with unequal widths and spacings are fabricated on the sample surface. The two outer sensors are used as heaters, and the middle sensor is used as a detector. By measuring the respective 3ω signals of the two heaters and the 2ω signal of the detector, and then solving the inverse problem based on the 3D FEM, the film and substrate thermal conductivity, along with the film-substrate TBR of the heterostructure are fitted. In virtue of the three-sensor 3ω-2ω method, two typical WBG semiconductor heterostructure samples, i.e., GaN on SiC (#SiC) and GaN on Si (#Si) with GaN layer thicknesses of ~2.3 μm, are measured. The thermal conductivity of the GaN films, the thermal conductivity of the SiC and Si substrates, and the TBR between the GaN and substrates are all consistent with the representative results in the literature, verifying the reliability and accuracy of the method. This method will provide a comprehensive solution for the demands of thermophysical property measurements of various solid heterostructures.

## Supplementary Material

See the Supplementary Material for the sensitivity analysis of other three kinds of heterostructures (GaN on Si, $Ga_2O_3$ on SiC, and $aAl_2O_3$ on Si) (Section S1), the measurement details of the sensors' dimensions (Section S2), the measurement details of the sensors' TCRs (Section S3), and the error analysis (Section S4).

## Acknowledgments

This work was financially supported by the National Natural Science Foundation of China (Grant No. 51825601, U20A20301).

## Data Availability

The data that support the findings of this study are available from the corresponding author upon reasonable request.

## References

1. Z. I. Alferov, "Nobel Lecture: The double heterostructure concept and its applications in physics, electronics, and technology," Reviews of Modern Physics **73**(3), 767 (2001).
2. H. Kressel, "Materials for Heterojunction Devices," Annual Review of Materials Science **10**(1), 287 (1980).
3. K. Yau, E. Dacquay, I. Sarkas, and S. P. Voinigescu, "Device and IC Characterization Above 100 GHz," IEEE Microwave Magazine **13**(1), 30 (2012).
4. S. J. Pearton and F. Ren, "GaN Electronics," Advanced Materials **12**(21), 1571 (2000).
5. A. J. Green, J. Speck, G. Xing, P. Moens, F. Allerstam, K. Gumaelius, T. Neyer, A. Arias-Purdue, V. Mehrotra, A. Kuramata, K. Sasaki, S. Watanabe, K. Koshi, J. Blevins, O. Bierwagen, S. Krishnamoorthy, K. Leedy, A. R. Arehart, A. T. Neal, S. Mou, S. A. Ringel, A. Kumar, A. Sharma, K. Ghosh, U. Singisetti, W. Li, K. Chabak, K. Liddy, A. Islam, S. Rajan, S. Graham, S. Choi, Z. Cheng, and M. Higashiwaki, "β-Gallium oxide power electronics," APL Materials **10**(2), 029201 (2022).
6. A. J. Heeger, "25th Anniversary Article: Bulk Heterojunction Solar Cells: Understanding the Mechanism of Operation," Advanced Materials **26**(1), 10 (2014).
7. J. Chen, W. Ouyang, W. Yang, J.-H. He, and X. Fang, "Recent Progress of Heterojunction Ultraviolet Photodetectors: Materials, Integrations, and Applications," Advanced Functional Materials **30**(16), 1909909 (2020).
8. D. M. Rowe, "Recent developments in thermoelectric materials," Applied Energy **24**(2), 139 (1986).
9. A. Bar-Cohen, J. D. Albrecht, and J. J. Maurer. "Near-Junction Thermal Management for Wide Bandgap Devices," in *2011 IEEE Compound Semiconductor Integrated Circuit Symposium (CSICS)* (Curran Associates, Waikoloa, Hawaii, USA, 2011).
10. Z. Cheng, S. Graham, H. Amano, and D. G. Cahill, "Perspective on thermal conductance across heterogeneously integrated interfaces for wide and ultrawide bandgap electronics," Applied Physics Letters **120**(3), 030501 (2022).
11. Y. Won, J. Cho, D. Agonafer, M. Asheghi, and K. E. Goodson. "Cooling Limits for GaN HEMT Technology," in *2013 IEEE Compound Semiconductor Integrated Circuit Symposium (CSICS)* (Curran Associates, Monterey, CA, USA, 2013).
12. Y. Won, J. Cho, D. Agonafer, M. Asheghi, and K. E. Goodson, "Fundamental Cooling Limits for High Power Density Gallium Nitride Electronics," IEEE Transactions on Components, Packaging and Manufacturing Technology **5**(6), 737 (2015).

13. R. J. Warzoha, A. A. Wilson, B. F. Donovan, N. Donmezer, A. Giri, P. E. Hopkins, S. Choi, D. Pahinkar, J. Shi, S. Graham, Z. Tian, and L. Ruppalt, "Applications and Impacts of Nanoscale Thermal Transport in Electronics Packaging," Journal of Electronic Packaging **143**(2), (2021).
14. P. Jiang, X. Qian, and R. Yang, "Tutorial: Time-domain thermoreflectance (TDTR) for thermal property characterization of bulk and thin film materials," Journal of Applied Physics **124**(16), 161103 (2018).
15. A. J. Schmidt, R. Cheaito, and M. Chiesa, "A frequency-domain thermoreflectance method for the characterization of thermal properties," Review of Scientific Instruments **80**(9), 094901 (2009).
16. Z. Cheng, F. Mu, T. You, W. Xu, J. Shi, M. E. Liao, Y. Wang, K. Huynh, T. Suga, M. S. Goorsky, X. Ou, and S. Graham, "Thermal Transport across Ion-Cut Monocrystalline β-$Ga_2O_3$ Thin Films and Bonded β-$Ga_2O_3$–SiC Interfaces," ACS Applied Materials & Interfaces **12**(40), 44943 (2020).
17. G. Chen, "Non-Fourier phonon heat conduction at the microscale and nanoscale," Nature Reviews Physics **3**(8), 555 (2021).
18. J. L. Braun, D. H. Olson, J. T. Gaskins, and P. E. Hopkins, "A steady-state thermoreflectance method to measure thermal conductivity," Review of Scientific Instruments **90**(2), 024905 (2019).
19. Y. Song, D. Shoemaker, J. H. Leach, C. McGray, H.-L. Huang, A. Bhattacharyya, Y. Zhang, C. U. Gonzalez-Valle, T. Hess, S. Zhukovsky, K. Ferri, R. M. Lavelle, C. Perez, D. W. Snyder, J.-P. Maria, B. Ramos-Alvarado, X. Wang, S. Krishnamoorthy, J. Hwang, B. M. Foley, and S. Choi, "$Ga_2O_3$-on-SiC Composite Wafer for Thermal Management of Ultrawide Bandgap Electronics," ACS Applied Materials & Interfaces **13**(34), 40817 (2021).
20. A. Sarua, H. Ji, K. P. Hilton, D. J. Wallis, M. J. Uren, T. Martin, and M. Kuball, "Thermal Boundary Resistance Between GaN and Substrate in AlGaN/GaN Electronic Devices," IEEE Transactions on Electron Devices **54**(12), 3152 (2007).
21. R. J. T. Simms, J. W. Pomeroy, M. J. Uren, T. Martin, and M. Kuball, "Channel Temperature Determination in High-Power AlGaN/GaN HFETs Using Electrical Methods and Raman Spectroscopy," IEEE Transactions on Electron Devices **55**(2), 478 (2008).
22. M. Kuball and J. W. Pomeroy, "A Review of Raman Thermography for Electronic and Opto-Electronic Device Measurement With Submicron Spatial and Nanosecond Temporal Resolution," IEEE Transactions on Device and Materials Reliability **16**(4), 667 (2016).
23. J. Cho, l. Zijian, M. Asheghi, and K. Goodson, "Near-Junction Thermal Management: Thermal Conduction in Gallium Nitride Composite Substrates," Annual Review of Heat Transfer **18**((2014).
24. Y.-C. Hua and B.-Y. Cao, "A two-sensor 3ω-2ω method for thermal boundary resistance measurement," Journal of Applied Physics **129**(12), 125107 (2021).
25. E. T. Swartz and R. O. Pohl, "Thermal resistance at interfaces," Applied Physics Letters **51**(26), 2200 (1987).
26. D. G. Cahill, "Thermal conductivity measurement from 30 to 750 K: the 3ω method," Review of Scientific Instruments **61**(2), 802 (1990).
27. D. G. Cahill and R. O. Pohl, "Thermal conductivity of amorphous solids above the plateau," Physical Review B **35**(8), 4067 (1987).
28. C. Dames, "Measuring the thermal conductivity of thin films: 3 omega and related electrothermal methods," Annual Review of Heat Transfer **16**((2012).
29. S. Deng, C. Xiao, J. Yuan, D. Ma, J. Li, N. Yang, and H. He, "Thermal boundary resistance measurement and analysis across SiC/$SiO_2$ interface," Applied Physics Letters **115**(10), 101603 (2019).

30. C. E. Raudzis, F. Schatz, and D. Wharam, "Extending the 3ω method for thin-film analysis to high frequencies," Journal of Applied Physics **93**(10), 6050 (2003).
31. Y. Liu, J. Qiu, L. Liu, and B. Cao, "Extracting optical constants of solid materials with micro-rough surfaces from ellipsometry without using effective medium approximation," Optics Express **27**(13), 17667 (2019).
32. C. Dames and G. Chen, "1ω, 2ω, and 3ω methods for measurements of thermal properties," Review of Scientific Instruments **76**(12), 124902 (2005).
33. B. W. Olson, S. Graham, and K. Chen, "A practical extension of the 3ω method to multilayer structures," Review of Scientific Instruments **76**(5), 053901 (2005).
34. A. Bedoya, J. Jaime, F. Rodríguez-Valdés, C. García-Segundo, A. Calderón, R. Ivanov, and E. Marín, "Thermal impedance," European Journal of Physics **42**(6), 065101 (2021).
35. M. Handwerg, R. Mitdank, Z. Galazka, and S. F. Fischer, "Temperature-dependent thermal conductivity and diffusivity of a Mg-doped insulating β-$Ga_2O_3$ single crystal along [100], [010] and [001]," Semiconductor Science and Technology **31**(12), 125006 (2016).
36. R. L. Xu, M. Muñoz Rojo, S. M. Islam, A. Sood, B. Vareskic, A. Katre, N. Mingo, K. E. Goodson, H. G. Xing, D. Jena, and E. Pop, "Thermal conductivity of crystalline AlN and the influence of atomic-scale defects," Journal of Applied Physics **126**(18), 185105 (2019).
37. D. Zhao, X. Qian, X. Gu, S. A. Jajja, and R. Yang, "Measurement Techniques for Thermal Conductivity and Interfacial Thermal Conductance of Bulk and Thin Film Materials," Journal of Electronic Packaging **138**(4), (2016).
38. A. T. Ramu and J. E. Bowers, "A "2-omega" technique for measuring anisotropy of thermal conductivity," Review of Scientific Instruments **83**(12), 124903 (2012).
39. E. Ziade, J. Yang, G. Brummer, D. Nothern, T. Moustakas, and A. J. Schmidt, "Thermal transport through GaN–SiC interfaces from 300 to 600 K," Applied Physics Letters **107**(9), 091605 (2015).
40. O. Madelung, U. Rössler, and M. Schulz (2002). "Silicon (Si), Debye temperature, heat capacity, density, hardness, melting point: Datasheet from Landolt-Börnstein - Group III Condensed Matter · Volume 41A1β," Springer-Verlag Berlin Heidelberg. https://doi.org/10.1007/10832182_478
41. P. Villars and F. Hulliger (2012). "$SiO_2$ ht1 heat capacity (specific heat): Datasheet from PAULING FILE Multinaries Edition – 2012," Springer-Verlag Berlin Heidelberg & Material Phases Data System (MPDS), Switzerland & National Institute for Materials Science (NIMS), Japan. https://materials.springer.com/isp/physical-property/docs/ppp_356e25e2153efbeebe626d9fdfef0a46
42. Y. C. Hua, H. L. Li, and B. Y. Cao, "Thermal Spreading Resistance in Ballistic-Diffusive Regime for GaN HEMTs," IEEE Transactions on Electron Devices **66**(8), 3296 (2019).
43. A. T. Ramu, N. I. Halaszynski, J. D. Peters, C. D. Meinhart, and J. E. Bowers, "An electrical probe of the phonon mean-free path spectrum," Scientific Reports **6**(1), 33571 (2016).
44. P. Jiang, X. Qian, X. Li, and R. Yang, "Three-dimensional anisotropic thermal conductivity tensor of single crystalline β-$Ga_2O_3$," Applied Physics Letters **113**(23), 232105 (2018).
45. J. Cho, Y. Li, W. E. Hoke, D. H. Altman, M. Asheghi, and K. E. Goodson, "Phonon scattering in strained transition layers for GaN heteroepitaxy," Physical Review B **89**(11), 115301 (2014).
46. J. Cho, Y. Li, D. H. Altman, W. E. Hoke, M. Asheghi, and K. E. Goodson. "Temperature Dependent Thermal Resistances at GaN-Substrate Interfaces in GaN Composite Substrates," in *2012 IEEE Compound Semiconductor Integrated Circuit Symposium (CSICS)* (Curran Associates, La Jolla, California, USA, 2012).

47. C. Y. Luo, H. Marchand, D. R. Clarke, and S. P. DenBaars, "Thermal conductivity of lateral epitaxial overgrown GaN films," Applied Physics Letters **75**(26), 4151 (1999).
48. T. L. Bougher, L. Yates, C.-F. Lo, W. Johnson, S. Graham, and B. A. Cola, "Thermal Boundary Resistance in GaN Films Measured by Time Domain Thermoreflectance with Robust Monte Carlo Uncertainty Estimation," Nanoscale and Microscale Thermophysical Engineering **20**(1), 22 (2016).
49. F. Mu, Z. Cheng, J. Shi, S. Shin, B. Xu, J. Shiomi, S. Graham, and T. Suga, "High Thermal Boundary Conductance across Bonded Heterogeneous GaN–SiC Interfaces," ACS Applied Materials & Interfaces **11**(36), 33428 (2019).
50. Y. Zhao, C. Zhu, S. Wang, J. Z. Tian, D. J. Yang, C. K. Chen, H. Cheng, and P. Hing, "Pulsed photothermal reflectance measurement of the thermal conductivity of sputtered aluminum nitride thin films," Journal of Applied Physics **96**(8), 4563 (2004).
51. A. Manoi, J. W. Pomeroy, N. Killat, and M. Kuball, "Benchmarking of Thermal Boundary Resistance in AlGaN/GaN HEMTs on SiC Substrates: Implications of the Nucleation Layer Microstructure," IEEE Electron Device Letters **31**(12), 1395 (2010).
52. Y.-H. Li, R.-S. Qi, R.-C. Shi, J.-N. Hu, Z.-T. Liu, Y.-W. Sun, M.-Q. Li, N. Li, C.-L. Song, L. Wang, Z.-B. Hao, Y. Luo, Q.-K. Xue, X.-C. Ma, and P. Gao, "Atomic-scale probing of heterointerface phonon bridges in nitride semiconductor," Proceedings of the National Academy of Sciences **119**(8), e2117027119 (2022).
53. J. Hu, Z. Hao, L. Niu, E. Yanxiong, L. Wang, and Y. Luo, "Atomically smooth and homogeneously N-polar AlN film grown on silicon by alumination of $Si_3N_4$," Applied Physics Letters **102**(14), 141913 (2013).
54. S. Zhang, Y. Zhang, Y. Cui, C. Freysoldt, J. Neugebauer, R. R. Lieten, J. S. Barnard, and C. J. Humphreys, "Interfacial Structure and Chemistry of GaN on Ge(111)," Physical Review Letters **111**(25), 256101 (2013).
55. A. Giri and P. E. Hopkins, "A Review of Experimental and Computational Advances in Thermal Boundary Conductance and Nanoscale Thermal Transport across Solid Interfaces," Advanced Functional Materials **30**(8), 1903857 (2020).